\documentclass[]{elsart}
\usepackage{natbib}
\usepackage{graphicx}
\usepackage{amsmath,amssymb,epsfig}
\renewcommand{\cite}{\citep}

\begin{document}
\begin{frontmatter}

\mathchardef\bsurd="1371
\mathchardef\bsolid="132E

\def\pd#1#2{\frac{\partial #1}{\partial #2}}
\def\bp{{\bf p}}
\def\bk{{\bf k}}
\def\bv{{\bf v}}
\def\bw{{\bf w}}
\def\bx{{\bf x}}
\def\br{{\bf r}}
\def\bz{{\bf s}}
\def\bb#1{{\bf #1}}
\def\be{{\bf e}}
\def\bn{{\bf n}}
\def\bq{{\bf q}}
%Poor man's bold
\def\pmb#1{\setbox0=\hbox{#1}%
\kern-.025em\copy0\kern-\wd0
\kern.05em\copy0\kern-\wd0
\kern-.025em\raise.0433em\box0}

%Bold operators and unusual symbols
\def\bze{{\pmb{$0$}}}
\def\bu{{\pmb{$1$}}}
\def\sm{{\pmb{$\cdot$}}}
\def\vm{{\pmb{$\times$}}}
\def\grad{{\pmb{$\nabla$}}}
\def\div{{\grad\sm}}
\def\curl{{\grad\vm}}

\def\bom{{\pmb{$\omega$}}}

\title{Vortex Splitting in Subcritical Nonlinear Schr\"odinger Equations}
\author{Natalia G. Berloff \corauthref{cor1}}
\corauth[cor1]{Email address: N.G.Berloff@damtp.cam.ac.uk}
\address{
Department of Applied Mathematics and Theoretical Physics\\
University of Cambridge,  Cambridge, CB3 0WA, United Kingdom
}
\begin{abstract}
Vortices and axisymmetric vortex rings are considered in the framework of the subcritical nonlinear Schr\"odinger equations. The higher order nonlinearity present in such systems models many-body interactions in superfluid systems and allows one to study the effects of negative pressure on  vortex dynamics. We find the critical  pressure for which the straight-line vortex becomes unstable to radial expansion of the core. The energy of the straight-line vortices and energy, impulse and velocity of vortex rings are calculated. The effect of a varying pressure on the vortex core is studied. It is shown that under the action of the periodically varying pressure field a vortex ring may split into  many vortex rings and the conditions for which this happens are elucidated. These processes are also relevant to experiments in Bose-Einstein condensates where the strength and the sign of two-body interactions can be changed via Feshbach resonance.
\end{abstract}
\begin{keyword}
 superfluidity, vortices, vortex rings,  nonlinear Schr\"odinger equation. 
\end{keyword}
%|
\end{frontmatter}

%==============================================
\section{Introduction} \label{sec:introduction}
%==============================================
The nonlinear Schr\"odinger (NLS) equation 
\begin{equation}
-i \psi_t = \frac{1}{2}\nabla^2 \psi + g|\psi|^2\psi,
\label{nls}
\end{equation}
where $g=\pm 1$, is one of the most studied equations of mathematical physics with applications ranging from nonlinear optics, where $\psi$ represents the electric field, Bose-Einstein condensates (BECs) and superfluidity, where $\psi$ is the field wavefunction, to fluid dynamics, where $\psi$ describes the amplitude of almost monochromatic wave. The nonlinear term in this equation represents the action of a refraction index that depends on the electric field intensity in nonlinear optics or corresponds to the two-body repulsive ($g=-1$) or attractive interactions ($g=1$) of the bosons in superfluids and BECs.

The formation of singularities in finite time for the focusing NLS equation (with $g=1$) is in practice been arrested by higher order nonlinear terms  that come from the expansion of the nonlinear refraction index in  optics, from many-body interactions in superfluids or from the effects of the geometry of the trapping potential in BECs that offers the reduction in the number of dimensions. These leads to a modification of the focusing NLS equation to what is known as ``subcritical'' or cubic-quintic NLS equation
\begin{equation}
-{\rm i} \psi_t = \frac{1}{2}\nabla^2 \psi + (g_1|\psi|^2-g_2 |\psi|^4)\psi.
\label{cq}
\end{equation}
This model was used to study cavitation and vortex nucleation (Josserand et al, 1995), formation of vapor droplets in superfluids (Josserand and Rica, 1997), solitary waves in nonlinear optics (Kivshar and Agrawal, 2002), and bright solitons in elongated BECs (Sinha et al, 2006; Khaykovich and Malomed, 2006).

In this paper we consider a  general class of subcritical NLS (SNLS) equations that allow for different orders of nonlinearity 
\begin{equation}
-{\rm i} \psi_t = \frac{1}{2}\nabla^2 \psi + (g_1|\psi|^2-g_2 |\psi|^{2(1+\gamma)})\psi,
\label{cq2}
\end{equation}
where $\gamma$ is an integer. The choice of the parameter $\gamma=1$ corresponds to the first-order term in the expansion of nonlinearity in the correlation energy and results in  the the cubic-quintic equation (\ref{cq}), whereas $\gamma=3$ is close to the value $2.8$ used in density-functional theories  (Dalfovo, 1992)  to produce a quantitatively correct equation of state for superfluid helium II.

Thus, we can view Eq. (\ref{cq2}) to be  a dimensionless form of a modified Gross-Pitaevskii equation
\begin{equation}
{\rm i} \hbar\psi_t = -\frac{\hbar^2}{2m}\nabla^2 \psi +\bigl(W_0|\psi|^{2(1+\gamma)}- V_0|\psi|^2\bigr)\psi,
\label{gp_gen}
\end{equation}
where $m$ is the mass of a boson, $V_0$ is a $\delta-$function two-body attractive interaction potential  and   many-body repulsive interactions are characterized by  a parameter $W_0$. Equation \ref{gp_gen} can be written in the Hamiltonian form 
\begin{equation}
{\rm i\hbar} \frac{\partial\psi}{\partial t} = \frac{\delta H}{\delta \psi^*},
\end{equation}
where the Hamiltonian (energy functional) of the system is
\begin{equation}
H=\int_V \biggl(\frac{\hbar^2}{2m}|\nabla\psi|^2 - \frac{V_0}{2}|\psi|^4 + \frac{W_0}{2 + \gamma}|\psi|^{2(2 + \gamma)}\biggr)\, d{\bf x}.
\label{h}
\end{equation}
To find the ground state the energy functional (\ref{h}) has to be minimized subject to the conservation of the number of particles $N=\int |\psi|^2\, d{\bf x}$. This can be achieved by introducing a Lagrange multiplier $\mu$ and minimizing $H-\mu N$. The energy of the ground state becomes
\begin{equation}
H_0= - \frac{V_0}{2}\frac{N^2}{V} + \frac{W_0}{2 + \gamma}\frac{N^{2 + \gamma}}{V^{1 + \gamma}},
\label{h0}
\end{equation}
and the ground state is given by $\psi_0^2=N/V$, $V=\int\, d{\bf x}$.
 The chemical potential  $\mu$ can be introduced into (\ref{gp_gen}) explicitly by
\begin{equation}
\psi\rightarrow \psi \exp(-{\rm i}  \mu t/\hbar),
\label{mu}
\end{equation}
so that (\ref{gp_gen}) becomes
\begin{equation}
{\rm i} \hbar\psi_t = -\frac{\hbar^2}{2m}\nabla^2 \psi +\bigl(W_0|\psi|^{2(1+\gamma)}- V_0|\psi|^2-\mu\bigr)\psi.
\label{gp_gen2}
\end{equation}
The ground state, $\psi_0,$ gives the value of the chemical potential as 
\begin{equation}
\mu = W_0\psi_0^{2(1+\gamma)}-V_0 \psi_0^2.
\end{equation}

The hydrostatic pressure of the system is found as 
\begin{equation}
P=-\frac{\partial H_0}{\partial V}=\frac{(1+\gamma) W_0}{2 +\gamma}\psi_0^{2(2+\gamma)} - \frac{V_0}{2}\psi_0^4.
\label{pressure}
\end{equation}
Hydrodynamic relation for the compressibility 
\begin{equation}
\frac{1}{m c^2}=\frac{\partial n}{\partial P}
\end{equation}
where $n=|\psi_0|^2$ is the number density, gives the expression for the speed of sound, $c$ as
\begin{equation}
c^2 = \frac{(1+\gamma) W_0 n^{1+\gamma} - V_0 n}{m}.
\label{c2}
\end{equation}
We conclude that the SNLS model, in particular, can be used to study the effects of a negative pressure. This is not possible with the cubic NLS  model (\ref{nls}) that is often used as a phenomenological model of superfluid helium, but where $P \sim \rho^2$.

In the last couple of decades there has been a number of experiments which explore the behaviour of superfluid helium at negative pressure by means of ultrasound waves that produce an oscillating pressure within a small volume of helium (Nissen et al, 1989; Maris and Xiong, 1989; Xiong and Maris, 1989).  In BECs experiments, Feshbach resonance is used to change the magnitude and sign of the scattering length. This can be modelled by Eq. (\ref{gp_gen}) with a periodically varying $V_0(t)$ in the presence of the external magnetic trap. In this paper we shall study the effects of variations of $V_0$ on the vortex structure and dynamics.

We cast Eq. (\ref{gp_gen2}) into dimensionless form by
\begin{equation}
{\bf x} \rightarrow b \, {\bf x}, \qquad t \rightarrow \frac{\hbar}{W_0 \psi_0^{2(1+\gamma)}}t, \qquad \psi\rightarrow \psi_0\psi,
\label{nond}
\end{equation}
where the healing length, $b$, is defined by 
\begin{equation}
b=\frac{\hbar}{\sqrt{W_0 m \psi_0^{2(1+\gamma)}}},
\label{b}
\end{equation}
so that Eq. (\ref{gp_gen2}) becomes
\begin{equation}
-i \psi_t = \frac{1}{2}\nabla^2 \psi +\bigl(|\psi|^{2(1+\gamma)}- 2 \xi|\psi|^2+(1 - 2 \xi)\bigr)\psi,
\label{main}
\end{equation}
where we denoted $\xi = V_0/2W_0\psi_0^\gamma$.

At $P=0$, $n_0=(V_0(2+\gamma)/2(1+\gamma)W_0)^{1/\gamma}$ and the speed of sound is $c_0^2=(1+\gamma)\gamma W_0 n_0^{1+\gamma}/(2+\gamma)$.The healing length becomes $b=\hbar/m c_0\sqrt{\gamma(1+\gamma)/(2+\gamma)}$. Since the known speed of sound in superfluid helium is approximately $238$ m s${}^{-1}$, the healing length at zero pressure is $b=0.6685 \sqrt{\gamma(1+\gamma)/(2+\gamma)}$ \AA, which for $\gamma=3$ gives $b=1$ \AA \, and the unit of time as  $6.28 \times 10^{-13}$s. 

In dimensionless units, the wavefunction in the bulk is $\psi_0=1$, the density is  $\rho=|\psi|^2$, and  the local pressure and the local speed of sound are
\begin{eqnarray}
P&=&\frac{1+\gamma}{2+\gamma}\rho^{2 + \gamma} - \xi \rho, \label{press}\\
c&=&\sqrt{(1+\gamma)\rho^{1+\gamma}-2 \xi\rho}.
\label{cc}
\end{eqnarray}
The paper is organized as follows. In Section 2 we study the amplitude and energy of the straight-line vortex at various pressures (interactomic strengths). We find the critical value of $\xi$ at which the vortex becomes unstable to radial expansion for various values of $\gamma$. Section 3 is devoted to vortex rings and other travelling coherent structures that propagate with a fixed velocity. We calculate the energy and impulse of these matter waves. The time evolution of a line vortex and a vortex ring with a periodically varying pressure is studied in Section 4. We show that the vortex ring splits  into many vortex rings if the negative pressure reaches the critical value for the vortex instability.
We conclude with Section 5.
%
%
%

%==============================================
\section{Vortex Lines} \label{sec:line}
\begin{figure}
\caption{The amplitudes of  straight-line vortices for $s=\pm 1$ as solutions of Eq. (\ref{vort}) for various $\gamma=1,2,3,4,5$ and $\xi=(1+\gamma)/(2+\gamma)$ that  correspond to zero pressure. The larger  values of $\gamma$ correspond to   tighter  vortex cores.}
\centering
\bigskip
\epsfig{figure=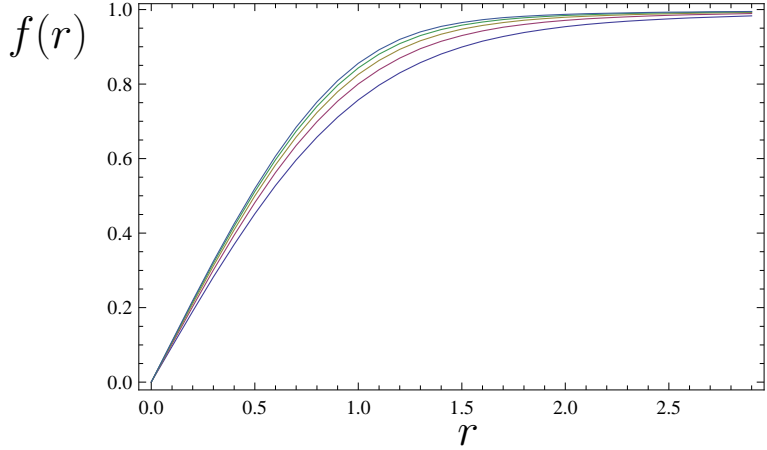, height=2.2 in}
\begin{picture}(0,0)
\put(-290,140) {\Large $f(r)$}
\put(-120,-10) {\Large $r$}
\end{picture}
\label{fig:vortexcore}
\end{figure}

\begin{figure}
\caption{The slopes at the origin, $a_1$, of the amplitudes of  straight-line vortices for $s=\pm 1$ as solutions of Eq. (\ref{vort}) for various $\gamma=1,2,3,4$ as functions of $\xi$. Dots represent numerical integration, solid lines -- Eq.(\ref{a1}). The values of $(\xi,a_1)$ that correspond to zero pressure are shown by red crosses.}
\centering
\bigskip
\epsfig{figure=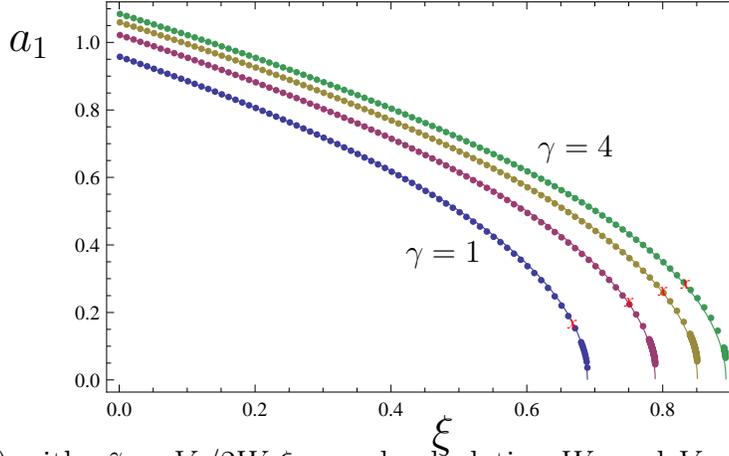, height=2.2 in}
\begin{picture}(0,0)
\put(-280,140) {\Large $a_1$}
\put(-120,-10) {\Large $\xi$}
\put(-130,60) {$\gamma=1$}
\put(-80,100) {$\gamma=4$}
\end{picture}
\label{fig:slopesG2}
\end{figure}

%==============================================
%
%
%
%
A vortex line is defined by a zero of the wave function $\psi = 0$.
In cylindrical coordinates $(r,\theta,z)$ the wave function of the straight-line
vortex takes form
\begin{equation}
\psi=f(r) \exp[i s \theta],
\end{equation}
where $s$ is an integer (``winding number'', ``topological charge'').
Fluid rotates around the $z$-axis with the tangential velocity
\begin{equation}
{\bf u}=s \nabla \theta= \frac{s}{r}{\bf e}_\theta
\end{equation}
 and the amplitude, $f$,  satisfies
 \begin{equation}
\frac{1}{2}\biggl[\frac{1}{r}\frac{d}{dr}\biggl(r \frac{d
 f}{dr}\biggr) - \frac{ s^2}{ r^2}f\biggr]+ 
 2 \xi f^3+(1-2\xi) f-f^{2(1+\gamma)+1}=0.
\label{vort}
\end{equation}
The boundary conditions are $f(0)=0$ and $f \rightarrow 1$ as $r\rightarrow\infty$. Analysis of Eq. (\ref{vort}) shows that near the origin $f(r)\sim a_s r^{|s|}$ and at infinity 
\begin{equation}
f(r)\approx 1-\frac{s^2}{4 (1 + \gamma - 2 \xi)r^2}-\frac{s^2 (2 \gamma^2 s^2 + \gamma (8 + 3 s^2) - (8 + s^2) ( 2 \xi-1))}{32 (1 + \gamma - 
   2 \xi)^3 r^4}+\cdot\cdot\cdot.
\end{equation}
Similar to the case of the cubic defocusing NLS equation, only vortices with single unit of quantization, $s=\pm1$, are dynamically stable (Josserand et al, 1995). Equation {\ref{vort}} was  solved  using a finite-difference
discretization  and Newton-Raphson iterations. Figure \ref{fig:vortexcore} gives plots of the vortex  amplitudes $f$
as  functions of  $r$. Figure \ref{fig:slopesG2} depicts the slopes at the origin of the vortex amplitudes, $a_1$, as functions of $\xi$ for various values of $\gamma$. Above a critical value of  $\xi_{\rm crit}(\gamma)$ the vortex becomes unstable and expands in a radial direction as seen by numerically integrating Eq. (\ref{main}) starting with the stable vortex and raising the value of $\xi$.  The existence of a critical pressure for the vortex stability has been established  in the context of  density-functional theory by  Xiong and Maris (1991) (giving the value of the critical pressure as $-6.5$ bars)  and by a more sophisticated theory by  Dalfovo (1992) (giving the value of $-8$ bars). We observe that the instability occurs when the slope of the vortex amplitude at the center of the vortex becomes zero and the negative pressure forces can no longer be balanced by the centrifugal energy of the fluid flow.

We can find the values of the critical pressure for the instability by evaluating Eq. (\ref{press}) with $\rho^\gamma=V_0/2W_0\xi_{\rm crit}$ and calculating $W_0$ and $V_0$ at zero pressure using the speed of sound $238$ m s${}^{-1}$, mass of the boson  $6.628 \times 10^{-27}$ kg and the density $145.2$ kg m${}^{-3}$. For $\gamma=3$ we obtained the critical pressure as $-6$ bar, that agrees quite well with the  estimates using much more sophisticated density-functional theories.

\medskip

Phenomenologically, the slope at the origin obeys a simple  law
\begin{equation}
a_1(\xi)=a_1(0)\sqrt{1-\xi/\xi_{\rm crit}},
\label{a1}
\end{equation}
where $a_1(0)=\lim_{\xi\rightarrow 0} a_1(\xi)$ and $\xi_{\rm crit}=a_1^{-1}(0)$. Equation (\ref{a1}) gives the slope at the origin accurate to two significant digits. For instance, our calculations agree with the result presented by Josserand et al (1995) done for $\gamma=1$. Numerically, we get $a_1=0.286$ for $\xi=5/8$, and Eq. (\ref{a1}) gives $a_1=0.29$. Table 1 shows the values of $a_1(0)$ and $\xi_{\rm crit}$ for various $\gamma$.

\centerline {\bf Table 1.} 
\medskip
\begin{tabular}{|c|cccc|}
	\hline
${}$ & $\gamma=1$ & $\gamma=2$ & $\gamma=3$ & $\gamma=4 $ \cr
\hline
$a_1(0)$ & 0.9575 &1.02155  & 1.05921 & 1.08409\\
$\xi_{\rm crit}$ &0.689 & 0.789 &  0.851 & 0.893\\
	\hline
\end{tabular}

To determine the energy of the vortex, we need to modify the energy functional (\ref{h}) by the presence of the ground state. To do this 
we restore dimensional units temporarily. Following Jones and
Roberts (1982) we denote by $\psi_u$  the wavefunction of the
undisturbed system of the same mass, so that 
\begin{equation}
\int\limits_V |\psi|^2\, d{\bf x} = \psi^2_uv,
\label{v}
\end{equation}
where
$v = \int_V \, d{\bf x}$,
  and by $\psi_\infty$ the wavefunction
of the bulk: $\psi \rightarrow \psi_\infty$ as $r\rightarrow \infty$.
For the quartic term in the expression for the energy we write using Eq. (\ref{v})
\begin{equation}
q_2=\int_V |\psi|^4 d{\bf x} - \psi_u^4 v=\int_V (|\psi|^2 - \psi_\infty^2)^2 d {\bf x}-(\psi_u^2-\psi_\infty^2)^2v.
\label{q2}
\end{equation}
Similarly, we obtained recursively
\begin{eqnarray}
q_3&=&\int_V |\psi|^6 d{\bf x} - \psi_u^6 v=\int_V (|\psi|^2 - \psi_\infty^2)^3d{\bf x}-(\psi_u^2-\psi_\infty^2)^3v+3\psi_\infty^2 q_2.\nonumber\\
q_{2+n}&=&\int_V |\psi|^{2(2+n)} d {\bf x} - \psi_u^{2(2+n)} v=\int_V (|\psi|^2 - \psi_\infty^2)^{2+n}d{\bf x}\nonumber \\
&&\hskip 40 pt-(\psi_u^2-\psi_\infty^2)^{2+n}v 
+\sum_{k=1}^{n}(-1)^{k-1}C_{2 + n}^k \psi_\infty^{2k} q_{2+n-k},
\label{qq}
\end{eqnarray}
where $n$ is a positive integer and $C_i^k$ are the binomial coefficients. The terms $(\psi_\infty^2 - \psi_u^2)^n v$ are $O(1/v)$ and vanish as $v\rightarrow \infty$.  In this limit, the dimensionless energy becomes
\begin{equation}
E_\gamma={1 \over 2 } \int|\nabla \psi|^2\, d {\bf x}
- \xi \int(|\psi^2|-1)^2\,d{\bf x}+ \frac{1}{2 + \gamma} Q_{2 + \gamma},
\label{ering}
\end{equation}
where $Q_{2+\gamma}$ are defined recursively as
\begin{eqnarray}
Q_2&=&\int(|\psi|^2 - 1)^3d{\bf x}, \nonumber \\
Q_{2+n}&=&\int (|\psi|^2 - 1)^{2+n}d{\bf x}+\sum_{k=1}^{n}(-1)^{k-1}C_{2 + n}^k  Q_{2+n-k}.
\label{qqq}
\end{eqnarray}

\begin{figure}
\caption{Values of vortex core parameter $\ell$  defined by (\ref{ev2}) of  straight-line vortices for $s=\pm 1$  for various $\gamma=1,2,3,4$ as functions of $\xi$. The values of the vortex core parameter $\ell$ at zero pressure are $-0.09$ for $\gamma=1$, $0.22$ for $\gamma=2$, $0.37$ for $\gamma=3$, $0.46$ for $\gamma=4$. The phenomenological fit (\ref{xiell}) is given by the solid lines.}
\centering
\bigskip
\epsfig{figure=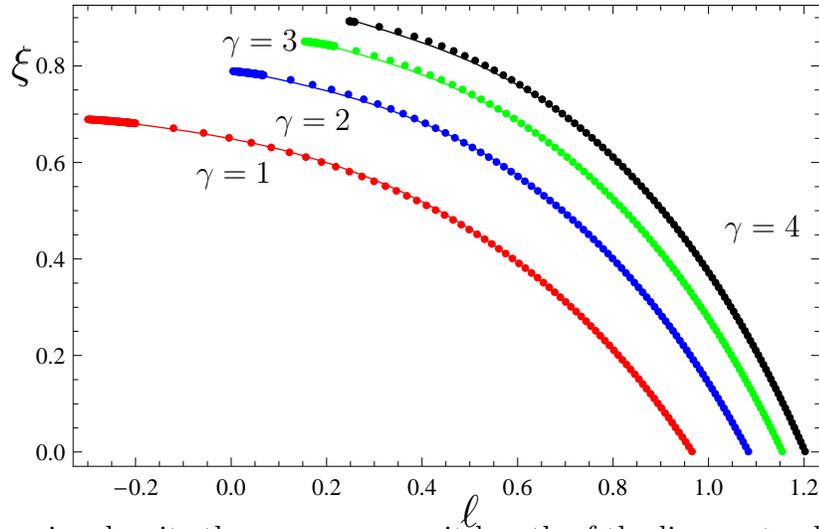, height=2.6 in}
\begin{picture}(0,0)
\put(-310,160) {\Large $\xi$}
\put(-140,-10) {\Large $\ell$}
\put(-240, 120) {$\gamma=1$}
\put(-210, 140) {$\gamma=2$}
\put(-230, 170) {$\gamma=3$}
\put(-40, 100)  {$\gamma=4$}
\end{picture}
\label{fig:ell}
\end{figure}

In dimensional units the energy per unit length of the line
 vortex becomes
\begin{eqnarray}
E_v &=& {\kappa^2 \rho_\infty \over 4 \pi} \biggl(\int\limits_0^\infty 
\left[\Bigl({d R \over dr}\Bigr)^2 + {R^2 \over r^2}\right]r\,dr
-2\xi\int_0^\infty(R^2-1)^2r\,dr\nonumber \\ \hskip 40 pt&+& \frac{2}{2+\gamma}\bigl[\int_0^\infty(R^2 - 1)^{2+\gamma}r\,dr\nonumber \\
\hskip 60 pt&+&
\sum_{k=1}^{\gamma}(-1)^{k-1}C_{2 + \gamma}^k \int_0^\infty (R^2 - 1)^{2+\gamma-k}r\,dr\bigr]\biggr).
\label{ev}
\end{eqnarray}
 The
second term in the first integral in Eq. (\ref{ev})  represents
the classical kinetic energy that diverges. This can be remedied
by introducing a cut-off distance $L$, corresponding to the
characteristic size of the container, and writing
\begin{equation}
\int_0^\infty\frac{R^2}{r}\, dr = \int_1^{L/b} \frac{1}{r} \,dr + \int_0^1\frac{R^2}{r}\, dr + \int_1^\infty\frac{R^2-1}{r}\, dr.
\end{equation}
The energy per
unit length of the line vortex can, therefore, be expressed in the form
\begin{equation}E_v = {\kappa^2 \rho_\infty \over 4 \pi} \left[\ln\Bigl({L\over b}\Bigr) + \ell\right],
\label{ev2}
\end{equation}
 where
 $\ell$ can be found by numerical integration.
Figure \ref{fig:ell} shows the values of the vortex core parameter $\ell$ as a function of $\xi$ for various values of $\gamma$.

Phenomenologically, the relationship between $\xi$ and $\ell$ has the form
\begin{equation}
\xi=\xi_{\rm crit} - \sigma_1(\ell- \ell_{\rm crit}) - \sigma_2(\ell- \ell_{\rm crit})^3,
\label{xiell}
\end{equation}
where $\ell_{\rm crit}$ is the value of $\ell$ at $\xi_{\rm crit}$. The values of parameters $\sigma_i$ and $\ell_{\rm crit}$ for various $\gamma$ are given in Table 2.

\centerline {{\bf Table 2.} }

The coefficients of Eq.(\ref{xiell}) that define the vortex core parameter $\ell$.

\medskip
\begin{tabular}{|c|cccc|}
	\hline
$$ & $\gamma=1$ & $\gamma=2$ & $\gamma=3$ & $\gamma=4$ \cr
\hline
$\ell_{\rm crit}$ & -0.299578 & 0.00410  & 0.15429 & 0.247549 \\
$\sigma_1$ &0.117906 & 0.20718 &  0.261971 &0.298006       \\
$\sigma_2$ &0.264568 & 0.449719 &  0.590682 &0.701111       \\
	\hline
\end{tabular}

\section{Vortex Rings}
In this section we consider the circular vortex rings that propagate in $z-$direction preserving their form. For large vortex rings with the radius $R$ much greater than the size of the core characterized by the healing length $b$ the energy can be found by using the energy of the straight line vortex (\ref{ev2})  in the region close to the vortex line and using only the kinetic energy term  in the region away from the vortex line.  The sum of these two contributions to energy gives a simple result (Amit and Gross, 1966; Roberts and Grant, 1971)
\begin{equation}
E = {1 \over 2} \kappa^2 \rho_\infty R\left[\ln\Bigl({8R \over b}\Bigr) 
- 2 + \ell \right].
\label{ering2}
\end{equation}
 The momentum  of the large vortex ring becomes
\begin{equation}
{\bf p}= \kappa \rho_\infty \pi R^2{\bf e_z},
\label{mring}
\end{equation}
where ${\bf e_z}$ is the unit vector in the direction of the ring motion.

The vortex rings and other travelling wave structures that propagate with a constant velocity correspond to stationary solutions of the SNLS equations in the frame of reference  moving with the velocity of the ring. The wavefunction satisfies 
\begin{equation}
{\rm i}U \frac{\partial \psi}{\partial  z} = \frac{1}{2}\nabla^2\psi+\bigl(|\psi|^{2(1+\gamma)}- 2 \xi|\psi|^2+(1 - 2 \xi)\bigr)\psi.
\label{umain}
\end{equation} 
We can perform a variation $\psi \rightarrow \psi  + \delta \psi$
in the expressions for momentum (see Jones and Roberts, 1982)
\begin{equation}
 {\bf p} = {1 \over 2{\rm i}}\int \left[(\psi^* - 1) {\bf\nabla} \psi - 
(\psi - 1){\bf\nabla} \psi^*\right]\, d {\bf x}
\label{pring}
\end{equation}
 and
energy (\ref{ering}) and using  (\ref{umain}) show that $\delta E = U \delta p$,
or
\begin{equation}
U=\frac{\partial E}{\partial p}.
\label{hgr}
\end{equation}
We can differentiate Eqs. (\ref{ering2}) and (\ref{mring}) with respect to $R$ and after substitution into Eq. (\ref{hgr}) obtain the expression for the velocity of the large vortex ring as 
\begin{equation}
 U = {\kappa \over 4 \pi R}\left[\ln\Bigl({8R \over b}\Bigr) - 1 + \ell\right].
\label{ularge}
\end{equation}
We can also use the Hamiltonian group relation (\ref{hgr}) to derive an alternative form of the energy functional (\ref{ering}). We substitute $z\rightarrow \alpha z$, for a constant $\alpha$ in Eqs\ (\ref{ering}) and (\ref{pring}). Then using
the variational relationship 
\begin{equation}
\frac{\partial}{\partial \alpha}\delta\left(E-Up\right)\bigg\vert_{\alpha=1}=0
\end{equation}
gives 
\begin{equation}
E=\int\left|\frac{\partial\psi}{\partial z}\right|^2\,d{\bf x}.
\label{e_alt}
\end{equation}
As the radius of the vortex ring decreases the expressions (\ref{ering2}), (\ref{mring}) and (\ref{ularge}) are no longer accurately describe the energy, impulse and velocity of the ring. The sequence of the small vortex rings and other localized disturbances can be found numerically by  a Newton-Raphson iteration technique. The infinite domain is mapped by the transformations $\widehat{x}=\tan^{-1}(Dx)$
and $\widehat{y}=\tan^{-1}(Dy)$ to a finite grid $(-\frac{\pi}{2},\frac{\pi}{2})\times(-\frac{\pi}{2},\frac{\pi}{2})$. $D$
is a constant chosen to lie in the range $D\sim 0.4-0.8$. The resulting equations are expressed in 
second-order finite-difference form. Taking $201^2$ grid points in the finite domain, the resulting non-linear equations are solved
by a Newton-Raphson iteration procedure using a banded matrix
linear solver based on the bi-conjugate gradient stabilized iterative method. The accuracy of the obtained solutions is verified by evaluating  the integral identities (\ref{hgr}) and (\ref{e_alt}).

\begin{figure}
\caption{Families of travelling wave solutions of the subcritical NLS equation (\ref{main}) for $\gamma=1$ (red thin solid line), $\gamma=2$ (blue thick solid line), and $\gamma=3$ (black dashed line) and  (a) $\xi=1/2$  and (b) $\xi=(1+\gamma)/(2+\gamma)$. Dots show the positions of the transition between a vortex ring and a rarefaction pulse on the lower branches. The critical velocities at which the vortex ring becomes the rarefaction pulse are given in Table 3.}
\centering
\bigskip
\vbox{\epsfig{figure=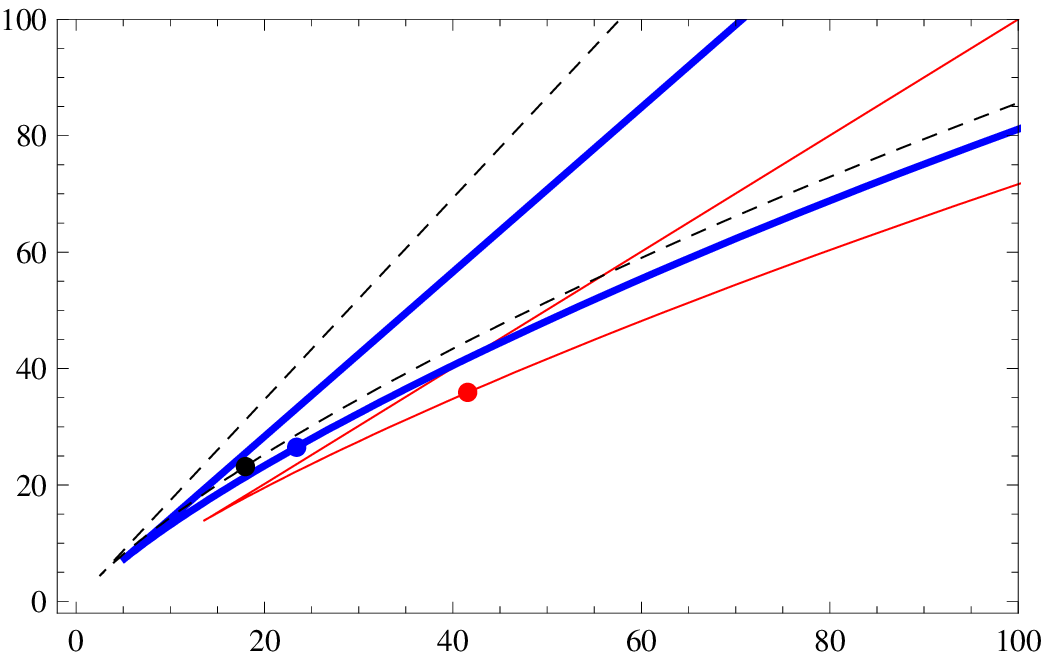, height=2.2 in}\\
\epsfig{figure=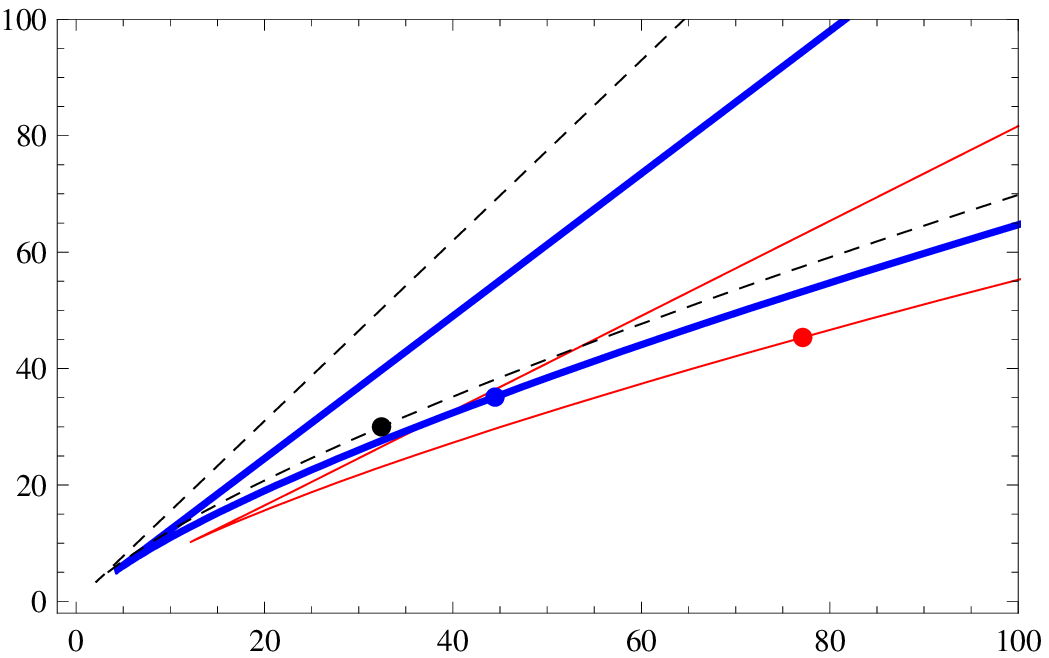, height=2.2 in}}
\begin{picture}(0,0)
\put(-150,300) { $E$}
\put(-100,270) { (a)}
\put(-150,140) { $E$}
\put(-100,120) { (b)}
\put(0,-0) { $p$}
\end{picture}
\label{fig:cusp}
\end{figure}

The families of the travelling wave solutions found are qualitatively similar to the sequence of such waves found by Jones and Roberts (1982) for the Gross-Pitaevskii model. They calculated the energy $E$ and momentum $p$
  and showed that  the sequence in the $p-E$ plane is given by two branches meeting at a cusp where $p$ and $E$
assume their minimum values,
$p_m$ and $E_m$. As $p \to \infty$
on each branch, $E \to \infty$. On the lower branch the
solutions were asymptotic to the large vortex rings. As $E$ and $p$ decrease from infinity along the
lower branch, the solutions begin to lose their similarity to
large vortex rings, and Eqs. (\ref{ering2}), (\ref{mring}) and (\ref{ularge}) determine $E$, $p$,
and $U$ less and less accurately.
Eventually, for a momentum $p_0$ slightly greater than $p_m$,
the rings lose their vorticity ($\psi$  loses its zero), and
thereafter the solitary solutions may better be described as
`rarefaction waves'. The upper branch consists entirely of these
and, as $p \to \infty$ on this branch, the solutions asymptotically
approach infinitesimal sound waves. Figure \ref{fig:cusp} shows the positions of such cusps for the SNLS equations (\ref{main}) for two values of $\xi$: $\xi=1/2$ and $\xi=(1+\gamma)/(2+\gamma)$.  The critical velocities at which the vortex ring becomes the rarefaction pulse are given in Table 3.

In Berloff and Roberts (1999) we discussed how a nonlocal  NLS equation with a higher order nonlinearity brings the vortex core parameter  $\ell$ and the healing length $b$ into the agreement. We observe that the same harmony is achieved by the subcritical NLS model (\ref{main}) for $\gamma=3$. At zero pressure that gives $b=1$ \AA, the same value as the one found from the sound speed in Section 1. This also agrees with the experiments of Rayfield and Reif (1964), where the vortex ring travelling at $27$ cm s${}^{-1}$ has the energy $10$ eV.

\centerline {{\bf Table 3.} }

The critical velocity of the transition between a vortex ring and a rarefaction pulse.

\medskip
\begin{tabular}{|c|ccc|}
	\hline
$\xi$ & $\gamma=1$ & $\gamma=2$ & $\gamma=3$  \cr
\hline
$\frac{1}{2}$ & 0.7 & 0.91  & 1.03 \\
$\frac{1+\gamma}{2+\gamma}$ &0.45 & 0.6 &  0.7        \\
	\hline
\end{tabular}

\section{Vortex splitting}
In this section we discuss a novel mechanism of creating vorticity in superfluids. Previously,  the semi-classical formation of vortices has been attributed  to one of the four basic mechanisms: due to the existence of critical velocities (Frisch et al, 1992), due to the transverse instabilities of dispersive waves (Kuznetsov and Rasmussen, 1995), due to collapse of cavities (Berloff and Barenghi, 2004), due to energy transfer between waves (Berloff, 2004) and during the condensation (Berloff and Svistunov, 2002). Here we show that there exists another mechanism of vortex formation based on the expansion and collapse of the vortex core.

A negative pressure is generated in superfluid helium II by an ultrasonic transducer that is used to produce periodic sound pulses (Maris, 1994). This technique has been extremely successful in detecting electron bubbles in superfluids  (Ghosh and Maris, 2005). It has been shown that the tensile strength of superfluid helium is much less than predicted by theory and this suggests that the quantized vortices may play a role in the cavitation process. 

To consider the dynamics of a straight line vortex and a vortex ring in superfluid under the time-varying pressure field, we integrated  Eq. (\ref{main}) numerically forward in time. We used the 4th order accurate  finite difference scheme in space and  the 4th order Runge-Kutta in time.

\begin{figure}
\caption{Density contour plots of the vortex under the action of  periodically varying pressure field characterized by Eq. (\ref{xii}) with $\epsilon=1/20$, $\gamma=1$ and $\eta=100$. The centre of the  vortex is seen as black region, intermediate densities are shown in light gray, larger densities in dark gray.}
\centering
\bigskip
\epsfig{figure=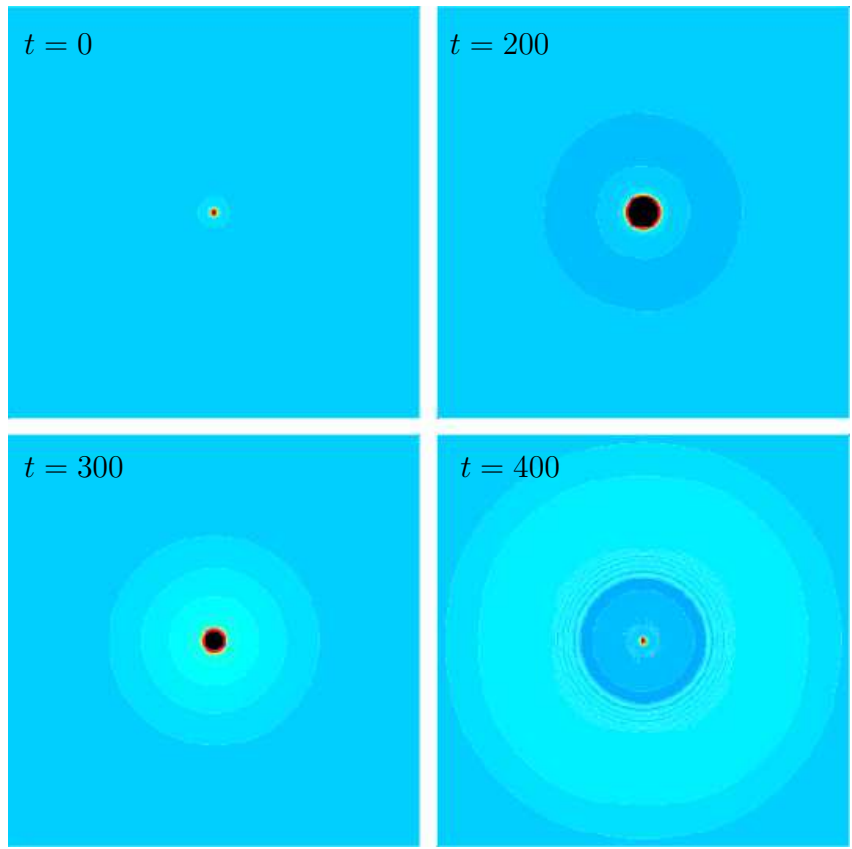, height=4.4 in}
\begin{picture}(0,0)
\put(-320,300) { $t=0$}
\put(-155,300) {$t=200$}
\put(-320,140) { $t=300$}
\put(-155,140) { $t=400$}
\end{picture}
\label{fig:strVort}
\end{figure}

We start by considering a single straight line vortex under the action of a varying pressure field controlled by parameter $\xi$. The pressure oscillates according to Eq. (\ref{press}) with 
\begin{equation}
\xi(t)=\xi_0+\epsilon \sin(\pi t/2\eta),
\label{xii}
\end{equation}
 where $\epsilon$ is a small parameter and  $\xi_0$ corresponds to  the initial zero pressure $\xi_0=(1+\gamma)/(2+\gamma)$. If $\xi_0+\epsilon\le \xi_{\rm crit}$, then the vortex  never becomes unstable and the size of the core increases until  $t=\eta$ and decreases back to the unperturbed value at $t=2\eta$ in response to the applied pressure. If $\xi_0+\epsilon > \xi_{\rm crit}$, then the vortex core continues to increase while pressure is negative. This instability becomes arrested by the growing positive pressure, so the vortex core returns to the unperturbed value at $t=4\eta$. Figure \ref{fig:strVort} shows the density contour plots for the dynamics of the vortex for $\epsilon=1/20$, $\gamma=1$ and $\eta=100$. The field around the vortex remains radially symmetric and there is no vortex splitting. The extra energy gained is emitted by the vortex core radially as sound waves. 

\begin{figure}
\caption{The vortex core parameter $\ell(t)$ in (\ref{ell_purt}) as a function of time for the dynamics of the vortex under the action of  periodically varying pressure field characterized by Eq. (\ref{xii}) with $\epsilon=1/20$ and  $\gamma=1$  (solid lines). Dashed line gives the evolution of pressure given by Eq.(\ref{press}) magnified by  a factor of  100 (top panel) or  500 (bottom panel).}
\centering
\bigskip
\vbox{\epsfig{figure=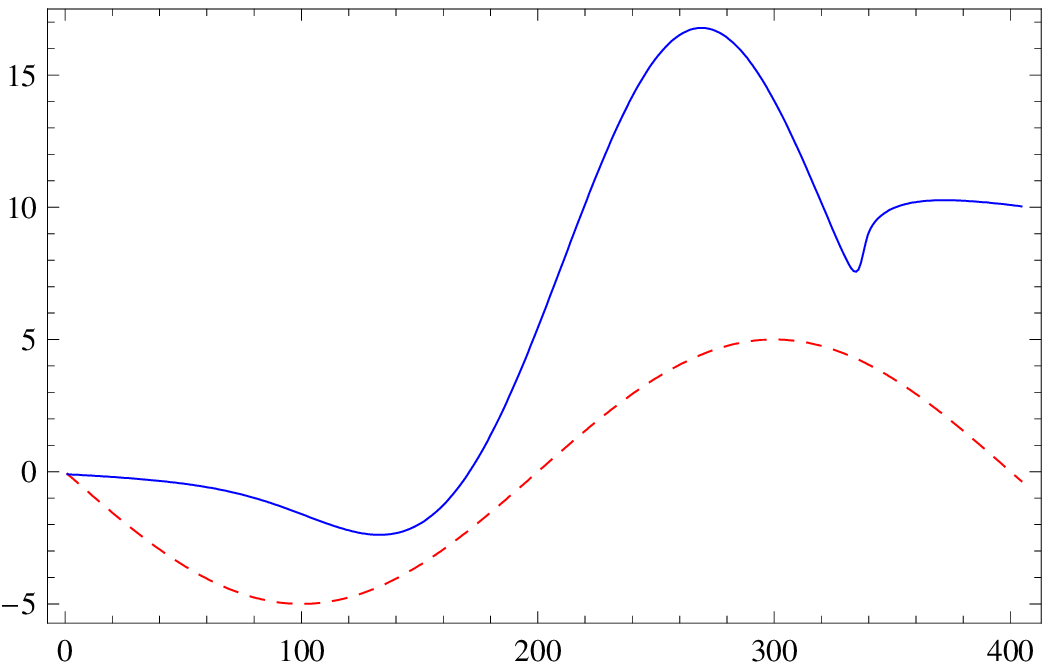, height=2.2 in}\\
\epsfig{figure=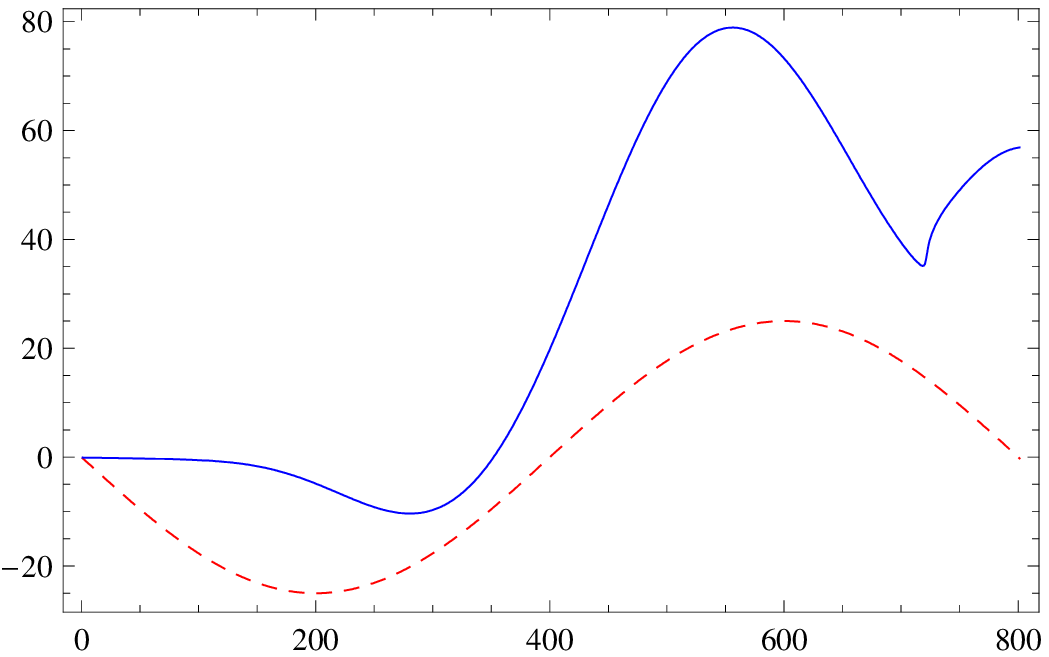, height=2.2 in}}
\begin{picture}(0,0)
\put(0,0) { $time$}
\put(-140,300) {$\ell(t)$}
\put(-140,140) { $\ell(t)$}
\put(-100,300) {$\eta=100$}
\put(-100,140) {$\eta=200$}
\end{picture}
\label{fig:pressEll}
\end{figure}

\begin{figure}
\caption{Time snapshots of the density of a straight-line vortex under the action of varying pressure field characterised by Eq. (\ref{xii}) with $\epsilon=1/20$. On the top panel the times are $0, 50,100,150,200,250,300$; on the bottom panel the times are $300,310,320,330,340,350$. The thinner lines correspond to later times.}
\centering
\bigskip
\vbox{\epsfig{figure=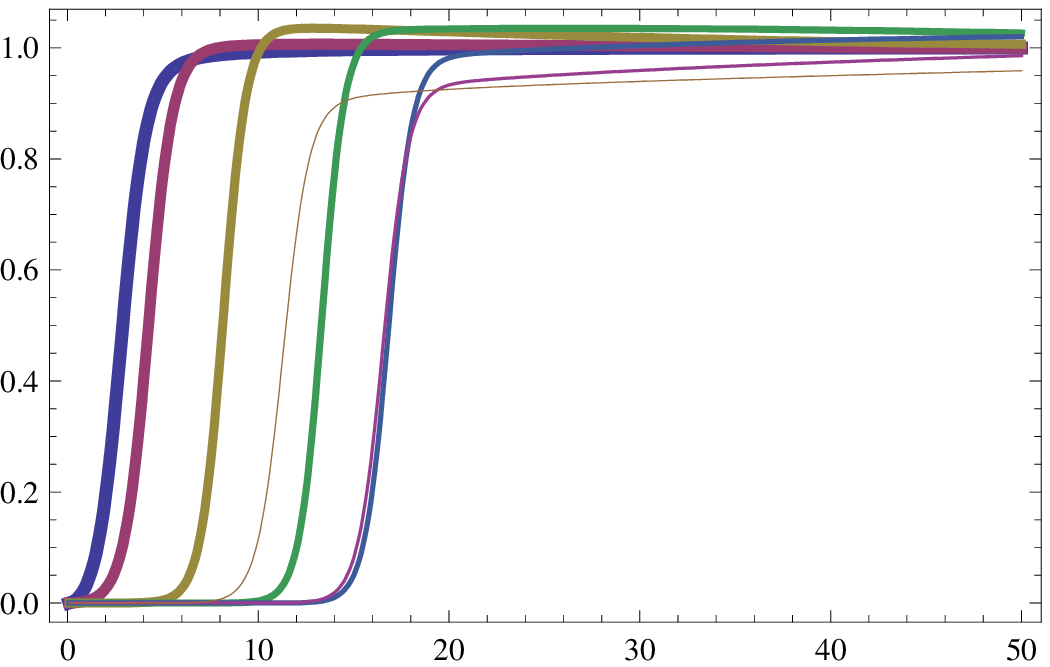, height=2.2 in}\\
\epsfig{figure=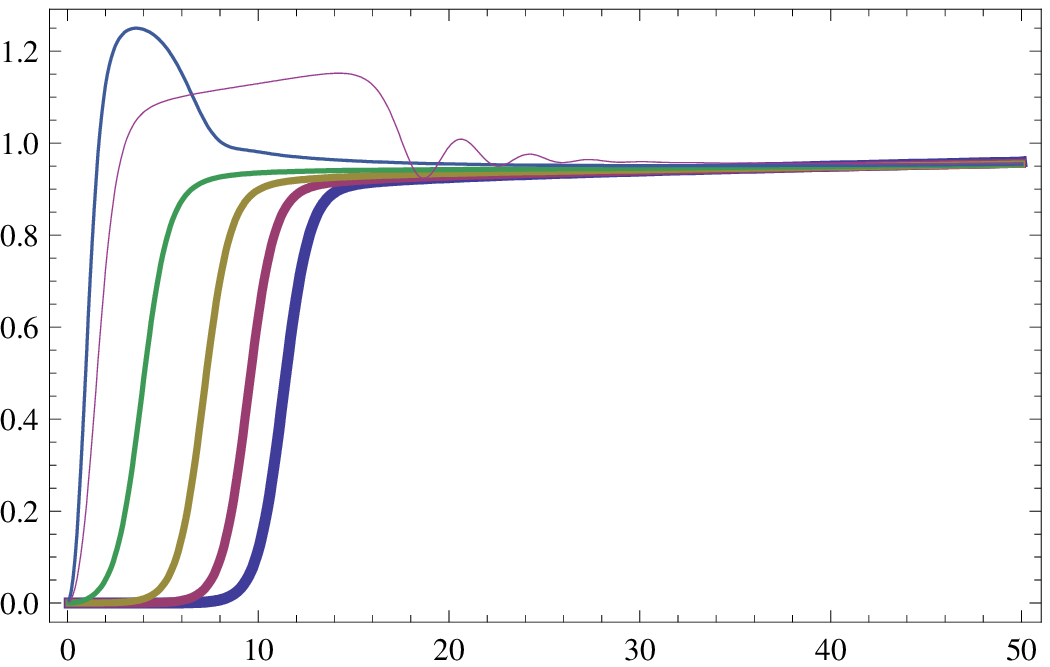, height=2.2 in}}
\begin{picture}(0,0)
\put(-140,300) {\Large $\rho$}
\put(0,0) {\Large $r$}
\end{picture}
\label{fig:vortCore}
\end{figure}

To understand the action of the pressure on the vortex we calculated the energy of the system  which in our dimensionless units is (see Eq. (\ref{ev2}))
\begin{equation}
E=\pi\biggl[\ln(L/b)+\ell(t)\biggr].
\label{ell_purt}
\end{equation}
Figure \ref{fig:pressEll} shows the plots of $\ell(t)$ as a function of $t$. The corresponding values of pressure are given by dashed line (and magnified by a factor of $100$ for $\eta=100$ and by a factor of $500$ for $\eta=200$.). The energy of the system, characterized by $\ell(t)$, increases between pressure minimum and maximum.  To follow the evolution of the vortex core we plot the density of the field in the radial direction on Fig. \ref{fig:vortCore} for $\epsilon =1/20$, $\gamma=1$ and $\eta=100$. The vortex core grows while the pressure stays negative (until $t=2\eta$), after that the core stabilizes and then  decreases in size as pressure increases; see the top panel of Fig. \ref{fig:vortCore}. For positive pressures the core continues to decrease in size while having an energy density much larger than the energy of the stationary vortex at a given positive pressure; a large density accumulates at the vortex core at $t=340$; see the bottom panel of Fig. \ref{fig:vortCore}. This extra energy is emitted by the vortex core as outgoing sound wave packet as it tries to regain its stationary density profile; see $t=350$ density profile at the bottom panel of Fig. \ref{fig:vortCore}. The increase in the period of the pressure oscillations leads to qualitatively similar scenario. The difference is that the energy of the vortex has a longer growth period and, therefore, reaches larger values; see the bottom panel of Fig. \ref{fig:pressEll}. 

\begin{figure}
\caption{Density contour plots of the cross section of the vortex ring under the action of  periodically varying pressure field characterized by (\ref{xii}) with $\epsilon=1/20$, $\eta=100$ and  $\gamma=1$. The cross-sections of vortex rings are seen as black dots, intermediate densities are shown in light gray, larger densities in dark gray. An initial vortex of radius $50b$ is split into three vortex rings at the end of one period of pressure oscillations.}
\centering
\bigskip
\epsfig{figure=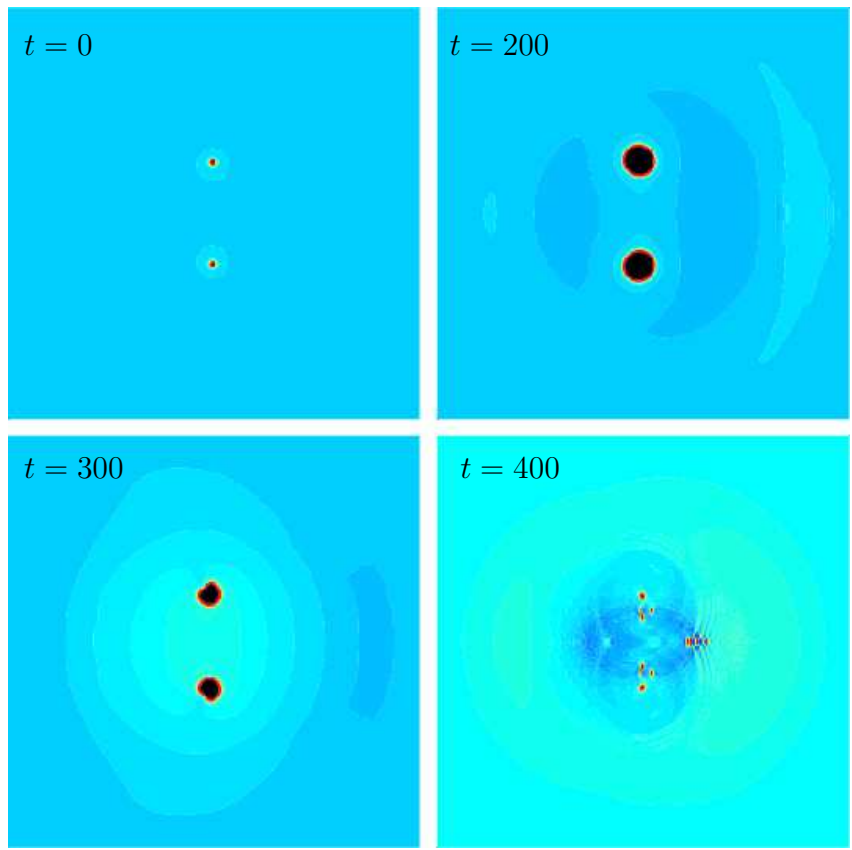, height=4.4 in}
\begin{picture}(0,0)
\put(-320,300) { $t=0$}
\put(-155,300) {$t=200$}
\put(-320,140) { $t=300$}
\put(-155,140) { $t=400$}
\end{picture}
\label{fig:vort1}
\end{figure}

\begin{figure}
\caption{Density contour plots of the cross section of the vortex ring under the action of  periodically varying pressure field characterized by Eq.(\ref{xii}) with $\epsilon=1/10$, $\gamma=3$ and $\eta=100$. The cross-sections of vortex rings are seen as black dots, intermediate densities are shown in light gray, larger densities in dark gray. An initial vortex of radius $50b$ is split into many vortex rings at the end of one period of pressure oscillations.}
\centering
\bigskip
\epsfig{figure=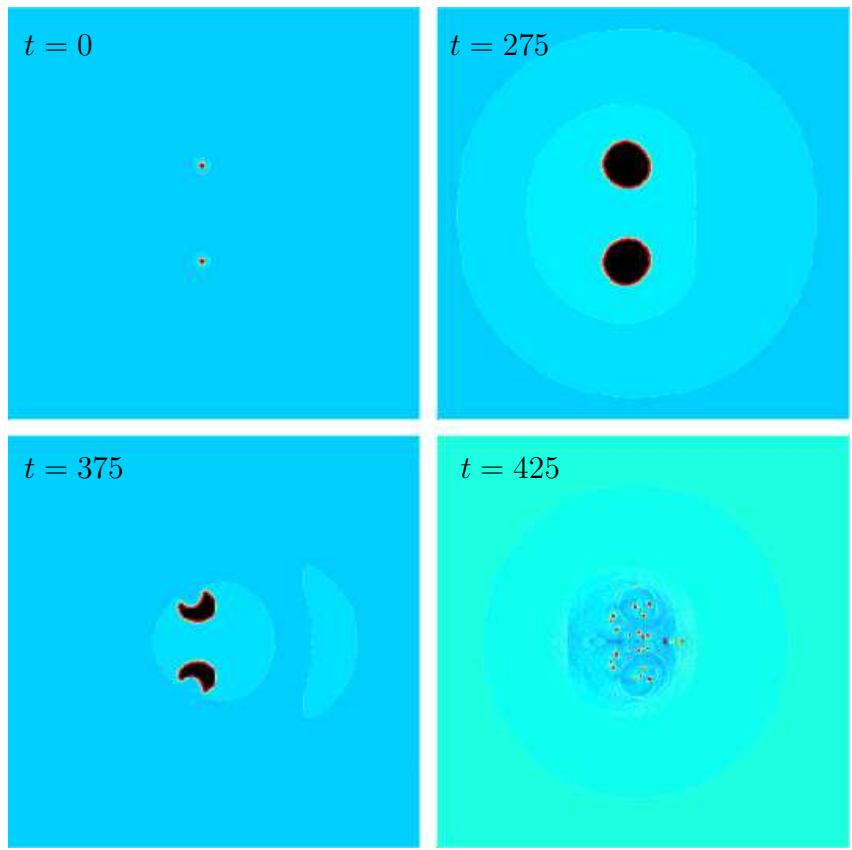, height=4.4 in}
\begin{picture}(0,0)
\put(-320,300) { $t=0$}
\put(-155,300) {$t=275$}
\put(-320,140) { $t=375$}
\put(-155,140) { $t=425$}
\end{picture}
\label{fig:vort3}
\end{figure}

For the vortices of large radii the effect of the pressure varying field is similar to that on the straight-line vortex, but the velocity field created by the vortex loop  changes the pressure distribution around the core. The vortex core collapses faster where the velocity is lower leading to the vortex splitting as Figs. \ref{fig:vort1} and \ref{fig:vort3} illustrate for $\gamma=1$ and $\gamma=3$ respectively.  In both cases we start with a vortex ring of radius $50$ at zero pressure. The parameter $\xi$ varies according to Eq. (\ref{xii}) with $\epsilon=1/20$ (for $\gamma=1$), $\epsilon=1/10$ (for $\gamma=3$)  and $\eta=100$. The vortex ring splits into three rings (two of positive circulation and one ring with the negative circulation) on Fig. \ref{fig:vort1}. There are many more vortex rings generated on Fig. \ref{fig:vort3}. We can roughly estimate the number of vortex rings that will be created by calculating the available energy just before splitting. Under the action of the pressure field characterized by (\ref{xii}) with $\epsilon=1/20$, $\gamma=1$ and $\eta=100$,  the single straight-line vortex emits sound when $\ell\approx 8$ (seen as a sharp cusp on Fig. (\ref{fig:pressEll})). The total energy available can be estimated from Eq. (\ref{ering}) and for $R=50$   is about 3 times the energy of one vortex ring of radius $50$ at zero pressure.
\section{Conclusions}
In summary, we considered the families of straight-line vortices and axisymmetric vortex rings in the framework of subcritical NLS equations with a focusing cubic nonlinearity and defocusing higher order term, characterized by the parameter $\gamma$; see Eq. (\ref{main}). These models can be used to study the effects of negative pressure in superfluids and capture the dynamics of the many body systems with attractive two-body interactions  such as spin-polarized Li${}_7$ (Bradley et al, 1995; Sackett et al, 1998). The strength of the two-body interactions as well as the hydrodynamic  pressure are characterized by the parameter $\xi$.

We determined the structure and energy of the straight-line vortices and showed that the vortex becomes unstable to uniform radial expansion when the slope of the amplitude at the vortex center becomes zero and  calculated the critical value of $\xi$ for this instability. The corresponding critical pressure for $\gamma=3$ is close to the one obtained by much more sophisticated density-functional theories. We calculated the energy, impulse and velocity of the axisymmetric vortex rings and showed the families of the travelling coherent structures on the impulse-energy plane. 

We considered the periodic oscillations of the parameter $\xi$ in time and elucidated its effect on the vortex core. There are three possible scenarios of the dynamics of the vortex. 
\begin{itemize}
\item[{1.}] If the value of $\xi$  never reaches the critical value for the vortex instability, then the vortex core simply adjusts to the varying pressure. As pressure decreases (increases), the vortex size grows (shrinks).

\item[{2.}] If the value of $\xi$ exceeds the critical value during the time evolution, then the vortex  core grows while pressure stays negative and continues to do so until shortly before pressure reaches its maximum positive value. After that the vortex core shrinks and the extra energy is emitted as outgoing sound waves.

\item[{3.}] Same as in item 2, but in the presence of large variations of the velocity field around the core, such as in the case of the vortex rings, the vortex core breaks into odd number of vortices (vortex rings) to preserve the total unit of circulation of $\pm 1$ around the initial vortex.
\end{itemize}

Finally, we suggest that the processes of the vortex instability and vortex  splitting can be seen in trapped BECs  where the magnitude and sign of the scattering length, which is represented by $\xi$ in our model, are changed by Feshbach resonance. In particular, our results indicate that there is a critical value of the interatomic attractive interactions for which a straight-line vortex becomes unstable. It seems likely that this instability can be arrested by the presence of a harmonic trapping potential. By periodically varying the sign and the magnitude of the two-body  interactions in the trapped BEC it may be possible to produce the vortex spitting  if the trap is not too close to being perfectly axisymmetric.
\section*{Acknowledgments}
The author gratefully acknowledges  the financial support from the EPSRC-UK and useful discussions with Humphrey Maris.

%==============================================
% Reference
%==============================================

\end{document}